\documentclass[12pt]{article}
\usepackage{epsfig}
\newlength{\dinwidth} \newlength{\dinmargin}
\def\a{\alpha} \def\b{\beta}  \def\d{\delta}
 \def\f{\phi} \def\g{\gamma} \def\h{\eta}
 \def\j{\psi}  \def\l{\lambda} \def\m{\mu}
\def\n{\nu} \def\o{\omega} \def\p{\pi}  
\def\s{\sigma}   \def\x{\xi} \def\z{\zeta}
\def\D{\Delta}  \def\G{\Gamma} 
\def\L{\Lambda}

\def\co{{\cal O}}

\def\samedirarrows{\raisebox{-.4ex}{\rlap{$\to$}} \raisebox{.4ex}{$\to$}} 
\def\oppdirarrows{\raisebox{-.4ex}{\rlap{$\to$}} \raisebox{.4ex}{$\gets$}} 

\def\Bar#1{\overline{#1}}                       

\def\beq{\begin{equation}} \def\eeq{\end{equation}}
\def\beqa{\begin{eqnarray}} \def\eeqa{\end{eqnarray}}

\def\beqx{\begin{displaymath}} \def\eeqx{\end{displaymath}}

\def\beql{\begin{eqnarray}} \def\eeql{\end{eqnarray}}
\def\NO{\nonumber}

\newcommand{\gwig}{\mbox{\,\raisebox{.3ex}
    {$>$}$\!\!\!\!\!$\raisebox{-.9ex}{$\sim$}}\,}    

\begin{document}
\setcounter{page}{0}
\thispagestyle{empty}
\rightline{NIKHEF 00-022}
\rightline{August 2000}
\vspace{20mm}

\begin{center}
{\Large\bf\sc
Soft Gluon Resummation for 
Polarized } \\[2ex]
{\Large\bf\sc
Deep-inelastic Production of Heavy Quarks}
\end{center}
\vspace{10mm}
\begin{center}
{\large 
Tim Oliver Eynck$^a$ and Sven-Olaf Moch$^b$}\\[1ex]
\vspace{5mm}
{\it NIKHEF Theory Group\\
        Kruislaan 409, 1098 SJ Amsterdam, The Netherlands} \\
\end{center}
\vspace{15mm}

\begin{abstract}
We present the threshold resummation for the cross section 
of heavy quark production in polarized deep-inelastic scattering 
to next-to-leading logarithmic accuracy and in single-particle 
inclusive kinematics. 
We expand our resummed result in $\alpha_s$ to next-to-leading 
and next-to-next-to-leading order and study the impact of these higher 
order corrections on the charm structure function $g_1^{\rm c}$ 
in the kinematical range accessible to the HERMES and COMPASS experiments.
\end{abstract}

\vspace{8.5cm}
{\small{
~~~~~$^a${\tt{teynck@nikhef.nl}}~~~~$^b${\tt{t66@nikhef.nl}}}}

\newpage
\section{Introduction}

Spin dependent deep-inelastic lepton-hadron scattering has received much interest during 
the past years, see Refs.\cite{Anselmino:1995gn,Cheng:1996ri,Cheng:2000xb,Lampe:1998eu} 
for reviews. 
Presently, next-to-leading order analyses of the experimental data for the structure 
function $g_1$ allow for the extraction of polarized parton distributions.
However, these are considerably less constrained than their unpolarized 
counterparts. 
This is particularly true for the gluon spin distribution 
$\D \f_{g/P}$, which is currently determined 
from scaling violations.
Numerous polarized hard scattering processes are sensitive to the gluon spin distribution, 
and the first moment of $\D \f_{g/P}$ 
contributes to the sum rule for the nucleon's spin content~\cite{Ellis:1974kp,unknown:1974su}.

Deep-inelastic leptoproduction of heavy quarks proceeds predominantly through the boson-gluon 
fusion process, and as such yields important information about the gluon distribution 
inside the proton.
Polarized leptoproduction of charm quarks is therefore promising for an 
experimental determination of $\D \f_{g/P}$. 
Presently, this measurement can be realized by the HERMES experiment \cite{hermescharm}.
The upcoming COMPASS experiment \cite{compass} is dedicated to measure $\D \f_{g/P}$ 
directly through the boson-gluon fusion process and subsequent polarized open charm production.
COMPASS will probe both the region of photoproduction and deep-inelastic scattering  
and extend the kinematical range accessible to the HERMES experiment.

In general, fixed target experiments, such as HERMES or COMPASS with centre-of-mass 
energies of $\sqrt{S} = 7.4 {\rm{GeV}}$ and $14.1 {\rm{GeV}}$ respectively, 
operate rather close to the threshold for charm pair-production.
In this region of phase space, the perturbatively calculable hard scattering 
cross section has potentially large higher order double-logarithmic 
threshold corrections.
Thus, a resummation of these Sudakov logarithms to all orders 
in perturbation theory is required in order to regain control over the 
perturbative expansion.

The purpose of this letter is to carry out this Sudakov resummation, 
which is done in section~\ref{crosssectionsection(sic!)}.
We work to next-to-leading logarithmic (NLL) accuracy 
(see Ref.\cite{Kidonakis:1999ze} for a recent review), 
and our final result provides an approximate sum of the complete perturbative expansion if, 
at each order, the Sudakov corrections dominate. 
Previously, studies of threshold resummation for 
polarized scattering processes have been performed 
for the polarized Drell-Yan cross section in Ref.\cite{Weber:1992wd}.
By re-expanding our resummed cross section in $\a_s$ 
to next-to-leading (NLO) and next-to-next-to-leading order (NNLO), 
we provide in section~\ref{finitesection} estimates of the size of the exact higher order
corrections\footnote{For the exact NLO corrections to polarized heavy quark photoproduction 
cf. Refs.\cite{Bojak:1998bd,Bojak:1998zm,Contogouris:2000en,Merebashvili:2000ya}.},
the calculation of which has not yet been completed~\cite{SinNPB,Sinprep}.
Our estimates are valid in the region dominated by threshold logarithms and 
we study their impact on the charm structure function $g_1^{\rm c}$.
In the kinematical region relevant to HERMES and COMPASS, we investigate 
the sensitivity to the gluon spin distribution 
$\D \f_{g/P}$ and the dependence on the factorization scale. 
Section~\ref{conclusions} contains our conclusions.

\section{Resummed differential cross section}
\label{crosssectionsection(sic!)}

We begin with the definition of the exact single-particle 
inclusive (1PI) kinematics. We study the reaction 
of lepton ($\rm l$) -- proton ($\rm P$) scattering, 
\beq
{\rm l}(l) + {\rm P}(p) \rightarrow {\rm l}(l-q) + {\rm{Q}}(p_1) + X[\overline{\rm{Q}}](p_2')\, ,
\label{epkin}
\eeq
where ${\rm{Q}}$ is a heavy quark
and $X[\overline{\rm{Q}}]$ denotes any allowed hadronic final state 
containing at least the heavy antiquark.
Neglecting electroweak radiative corrections\footnote{ 
QED radiative corrections have been studied 
in the leading-logarithmic approximation in Ref.\cite{Schienbein:1997sb}.},
for longitudinally polarized protons, the difference 
of cross sections for reaction (\ref{epkin}) 
may after summation over $X$ be written 
as~\cite{SinNPB,Sinprep,Zijlstra:1994sh,Zijlstra:Erratum,Buza:1997xr}
\beql
\label{diffpolcross}
\lefteqn{
\frac{d^5\sigma^{\oppdirarrows}}{dx\,dy\,d\f\,dT_1\,dU_1} 
- \frac{d^5\sigma^{\samedirarrows}}{dx\,dy\,d\f\,dT_1\,dU_1} \,=}\\[1ex]
& &\frac{4 \alpha^2}{Q^2}\Bigg[
\Big(2-y-\frac{2M^2x^2y^2}{Q^2}\Big) \frac{d^2g_1(x,Q^2,m^2,T_1,U_1)}{dT_1\,dU_1}
-\frac{4M^2x^2y^2}{Q^2} \frac{d^2g_2(x,Q^2,m^2,T_1,U_1)}{dT_1\,dU_1}\Bigg]\,,
\nonumber
\eeql
where $\alpha$ and $M$ are the fine structure constant and the proton mass, and 
$\f$ denotes the angle between the proton spin and the momentum 
of the scattered lepton, $(l-q)$ in Eq.(\ref{epkin}). 
The lower arrow on $d^5\s$ indicates the polarization of the incoming lepton 
in the direction of its momentum $l$, whereas the upper arrow on $d^5\s$ 
represents the polarization of the proton, which is parallel or anti-parallel 
to the polarization of the incoming lepton.

The functions $d^2g_{k}/dT_1\,dU_1,\; k=1,2$ are the double-differential
deep-inelastic polarized heavy quark structure functions.
The kinematic variables $Q^2, x, y$ are defined by 
\beq
Q^2 = -q^2 > 0\quad,\quad x=\frac{Q^2}{2p\cdot q}
\quad,\quad y=\frac{p\cdot q}{p\cdot l}\, .
\eeq
Additionally, we also have the overall invariants
\beq
\label{hadronicinv}
S = (p+q)^2 \equiv S'-Q^2 ,\quad
T_1 = (p-p_1)^2-m^2 ,\quad
U_1 = (q-p_1)^2-m^2 ,\quad
S_4 = S+T_1+U_1+Q^2 .
\eeq

Only the structure function $g_{1}$ is relevant asymptotically for $M/Q \ll 1$. 
It contains leading twist two operators and has a parton model interpretation.
We therefore restrict ourselves to $g_{1}$ only. Furthermore, we neglect the 
contributions from initial state quarks and antiquarks.  
This is justified as they vanish at leading order.
We consider only the gluon-initiated partonic subprocess 
\beql
\g^*(q) +\, g(k) &\longrightarrow& 
{\rm{Q}}(p_1) +\, X'[{\Bar{\rm{Q}}}](p_2')\, ,
\label{photon-gluon-fusion-0}
\eeql
where $k = z\, p$. The partonic invariants are 
\beq
s=(k+q)^2 \equiv s'+q^2 , \ \ \ \ t_1=(k-p_1)^2-m^2 ,
\ \ \ \ u_1=(q-p_1)^2-m^2 , \ \ \ \ s_4=(p'_2)^2-m^2 .
\label{mandelstam-def1}
\eeq 
The invariant $s_4$ parametrizes the inelasticity of the partonic reaction 
(\ref{photon-gluon-fusion-0}) in 1PI kinematics. 
This distance from threshold is conveniently measured in terms of 
a dimensionless weight function $w_S$ \cite{Contopanagos:1997nh},
\begin{equation}
\label{eq:1piweight}
w_S = {s_4\over m^2} \simeq {2 \bar{p}_2\cdot k_S \over m^2} 
\equiv {2 \zeta \cdot k_S \over m}\, .
\end{equation}
The 1PI kinematics vector $\zeta$ is defined by splitting the recoil momentum $p_2' = \bar{p}_2 +k_S$ 
into the momentum $\bar{p}_2$ at threshold, and the momentum  $k_S$ of any additional (soft) radiation 
above threshold.
 
In double-differential form, $d^2g_1/dT_1\,dU_1$ factorizes as
\beql
S'^2 \frac{d^2g_1(x,Q^2,m^2,T_1,U_1)}{dT_1\,dU_1}
&=& \int\limits_{z^-}^{1}\frac{dz}{z}
\,\D \phi_{g/P}(z,\mu^2)\;
\D \omega_{g}\Big({x\over z},s_4,t_1,u_1,Q^2,m^2,\m^2,\alpha_s(\mu)\Big)\, .\,\,\,
\label{d2g1fact}
\eeql
The polarized gluon distribution in the proton 
is denoted by $\D \phi_{g/P}$ and $z$ is the momentum fraction 
with $z^- = -U_1/(S'+T_1)$. 
The dimensionless function $\D \o_{g}$ describes the hard 
scattering process and depends on the partonic invariants 
$t_1,u_1$ and $s_4$, defined in Eq.(\ref{mandelstam-def1}).
The factorization scale is denoted by $\mu$ and throughout this paper  
taken equal to the renormalization scale.

Sufficiently close to threshold, $g_1$ in Eq.(\ref{d2g1fact}) is dominated 
by those higher order contributions that contain plus-distributions 
of the type
\beq
\left[{\ln^{l}(s_4/m^2)\over s_4}\right]_+
= \lim_{\Delta \rightarrow 0} \Bigg\{
{\ln^{l}(s_4/m^2)\over s_4} \theta(s_4 -\Delta)
+ \frac{1}{l+1}\ln^{l+1}\Big({\Delta\over m^2}\Big)\, \delta({s_4}) \Bigg\}
\label{s4distdef}\,.
\eeq
They result from imperfect cancellations between soft and virtual contributions 
to the cross section. 
At order $\co(\alpha^{i+2}_s),\; i=0,1,\ldots$ we refer to the corrections with $l=2i+1$ as  
leading logarithmic (LL),  
to the ones with $l=2i$ as next-to-leading logarithmic (NLL), etc.  

The Sudakov resummation of these singular functions in Eq.(\ref{s4distdef}) uses 
the methods and results of Refs.\cite{Contopanagos:1997nh,Sterman:1987aj,Kidonakis:1997gm,Laenen:1998qw}.
It rests upon the factorization of $\D \omega_g$ into separate functions $\D \psi_{g/g}$, $S$, 
and $H_g$ for the jetlike-, soft, and off-shell quanta.
This factorization, valid in the threshold region of phase space, 
implies a decomposition \cite{Laenen:1998qw} of the total weight function $w$, 
\begin{eqnarray}
  \label{eq:kindef}
w\,=\, w_\j + w_S \,=\, w_1\left(\frac{-u_1}{m^2}\right) + w_S\, ,
\end{eqnarray}
which expresses the overall weight in terms of the individual weights $w_\j$ and $w_S$.
Each of the functions $\D \psi_{g/g}(w_1)$, $S(w_S)$, and $H_g$, which are 
conveniently computed in $\zeta \cdot A = 0$ gauge, organizes  
contributions corresponding to a particular set of quanta 
and thereby only depends on its own individual weight function.

It is convenient to consider these functions in moment space, 
defined by the Laplace transform with respect to the overall weight $w$,
\begin{equation} 
{\tilde f}(N) = \int\limits_0^\infty dw
\,e^{-N w} f(w)\,.
\label{laplacetfm}
\end{equation}

Under the assumption of factorization the partonic cross section $\D \omega_g$ 
in moment space can then be written in a factorized form, up to
corrections of order $1/N$, as \cite{Contopanagos:1997nh,Kidonakis:1997gm,Laenen:1998qw,Laenen:1998kp}
\beql
\label{threshmoment}
\lefteqn{
\D \tilde{\omega}_{g}(N,t_1,u_1,Q^2,m^2,\mu^2,\a_s(\mu)) \,=} \\
& &
H_{g}(\zeta,Q^2,m^2,\a_s(\m))
\left[ { \D {\tilde\psi}_{g/g}(N_u,k\cdot\zeta/\mu)
\over {\D \tilde\phi}_{g/g}(N_u,\mu)} \right] {\tilde S}\left({m\over N\mu},\zeta\right)\, .
\nonumber
\eeql
where $N_u = N (-u_1/m^2)$. The $N$-dependence in each of the functions 
of Eq.(\ref{threshmoment}) exponentiates.     

The jet and soft functions, $\D \psi_{g/g}$ and $S$ in Eq.~(\ref{threshmoment}), 
can each be represented 
as operator matrix elements~\cite{Sterman:1987aj,Kidonakis:1997gm,Laenen:1998qw,Laenen:1998kp}. 
Dependence on the spin degrees of freedom can be kept explicit in these 
elements, such that the methods of 
Refs.\cite{Contopanagos:1997nh,Sterman:1987aj,Kidonakis:1997gm,Laenen:1998qw} 
for the resummation of the singular functions in Eq.(\ref{s4distdef}) via appropriate
evolution equations can be generalized in a straightforward way to account for initial state 
polarizations.
The jet-function $\D {\tilde\psi}_{g/g}$ obeys two evolution equations, namely 
the renormalization group equation and an equation describing the energy dependence
via gauge-dependence \cite{Contopanagos:1997nh,Sterman:1987aj,Collins:1981uk}.
Solving both equations sums Sudakov double logarithms.
Mass factorization introduces the density $\D {\tilde\f}_{g/g}$ as a counterterm 
and requires the choice of a scheme. 
In the $\Bar{{\rm{MS}}}$-scheme $\D {\tilde\f}_{g/g}$ has no double logarithms.
The soft function $S$ obeys a renormalization group equation, which resums all NLL 
logarithms originating from soft gluons not already accounted for
in $\D {\tilde\psi}_{g/g}$. 
In general, the soft and the hard function $S$ and $H_g$ depend also on the colour 
structure of the underlying scattering reaction, but for deep-inelastic production of 
heavy quarks this dependence is trivial~\cite{Laenen:1998kp}.

The final result for the hard scattering function $\D \tilde{\o}_{g}$ in the
$\Bar{{\rm{MS}}}$-scheme
in moment space resums all $\ln N$ 
in single-particle inclusive kinematics. 
We obtain 
\beql
\label{deltaomegaresummed}
\D \tilde{\omega}_{g}^{\rm res}(N,t_1,u_1,Q^2,m^2,\mu^2,\a_s(\mu))
&=& \;\\
&\ & \hspace{-60mm} \;
H_{g}(\zeta,Q^2,m^2,\a_s(m))\;
{\tilde S}(1,\z) \exp\Bigg \{ 2\int\limits_{\mu}^{m}{d\mu'\over\mu'}
\gamma_g\left(\alpha_s(\m^{\prime})\right)\Bigg\}
\nonumber \\
&\ & \hspace{-60mm} \times\;
\exp \Bigg \{
\int\limits_0^{\infty}\! \frac{d w}{w}
\! \left(1 - {\rm{e}}^{-N_uw} \right)\!
\Bigl[\, \int\limits_{w^2}^1 \frac{d \l}{\l}
A^{g}(\a_s(\sqrt{\l} 2k\cdot \z)) + \frac{1}{2} \n_{g}(\a_s(w 2k\cdot \z)) \Bigr]
\Bigg \}
\nonumber\\
&\ & \hspace{-60mm}\times\;
\exp\Bigg \{\int\limits_{m}^{m/N}{d\mu'\over\mu'}
2\, {\rm Re} \Gamma_S\left(\alpha_s(\m^{\prime})\right) \Bigg\}
\exp\Bigg \{ -2\int\limits_{\mu}^{2k\cdot \z}{d\mu'\over\mu'}
\left(\gamma_g\left(\alpha_s(\m^{\prime})\right)-
\gamma_{g/g}\Big( N_u,\alpha_s(\m^{\prime})\Big)\right)
\Bigg\}\, ,\nonumber
\eeql    
where $N_u$ is defined below Eq.(\ref{threshmoment}).
To NLL accuracy, as defined below Eq.(\ref{s4distdef}), the product $H_{g} S$ on the second line 
of Eq.(\ref{deltaomegaresummed}) is determined from matching 
to the Born result at the scale $\m = m$.       
To this accuracy the product $H_{g} S$ is also insensitive to the choice of 
treatment of $\g_5$. 

The second exponent in Eq.(\ref{deltaomegaresummed}) gives the leading 
$N$-dependence of the ratio $\D{\tilde{\j}}_{g/g}/\D{\tilde{\f}}_{g/g}$ with 
\beql
\label{poljetfunc}
A_{g}(\a_s) \,=\, C_A \frac{\a_s}{\p} 
+ \frac{1}{2} C_A K \left(\frac{\a_s}{\p}\right)^2 + \dots \, ,
\quad\quad\quad 
\n^{g}(\a_s) \,=\, 2 C_A \frac{\a_s}{\p}+ \dots \, ,
\eeql
where $K=C_A(67/18-\p^2/6)-5/9n_f$ can be inferred from 
Refs.\cite{Mertig:1996ny,Vogelsang:1996im} using Ref.\cite{Kodaira:1982nh} 
(cf. also Ref.\cite{Weber:1992wd}). 
The scale evolution of the ratio
$\D{\tilde{\j}}_{g/g}/\D{\tilde{\f}}_{g/g}$ is governed by 
\beql
\label{polaltpar1}
\g_{g}(\a_s) &=& b_2 \frac{\a_s}{\p} + \dots\, , \\
\label{polaltpar2}
\g_{g/g}(N,\a_s) &=& - \frac{\a_s}{\p}\!\left(C_A \ln N  - b_2 \right)\!
- \!\left(\frac{\a_s}{\p}\right)^2\!\!\left(\frac{1}{2} C_A K \ln N \right)\!
+ \dots\, , 
\eeql
with $\g_{g/g}$ calculated in Refs.\cite{Mertig:1996ny,Vogelsang:1996im}, 
and the soft anomalous dimension to order $\a_s$ is
\beql
\G_S(\a_s)&=&\frac{\a_s}{\p}
\Biggl\{ \!\left(\frac{C_A}{2} - C_F\right)\! ( L_\b + 1 ) -
\frac{C_A}{2} \left( \ln \left({4 (k \cdot \z)^2\over m^2}\right) +
\ln\frac{m^4}{t_1\, u_1} \right)\! \Biggr\}+ \dots\, ,
\label{softadim-res}
\eeql
with $\b = \sqrt{1- 4\, m^2/s}$ and
$L_\b=(1-2\,m^2/ s)\{ \ln (1-\b)/(1+\b) + {\rm{i}}\p \}/\b$. 
We note that the various anomalous dimensions 
in Eqs.(\ref{poljetfunc})--(\ref{softadim-res}) are identical to NLL accuracy 
to the corresponding quantities of soft gluon resummation for 
unpolarized scattering~\cite{Laenen:1998kp}.

Eq.(\ref{deltaomegaresummed}) represents the central result of this paper.
It provides the sum of Sudakov logarithms due to soft gluon emission to all 
orders in the perturbative expansion and accurate to the next-to-leading 
logarithm.

\section{Finite order results}
\label{finitesection}

We expand our resummed result in $\a_s$ up to second order so as to provide 
NLO and NNLO approximations to NLL accurracy for the partonic 
single-particle inclusive double-differential cross section difference $\D \sigma_{\g^*g}$.

For the process $\g^* + g \longrightarrow
{\rm{Q}} + \Bar{\rm{Q}}$, the single-particle inclusive double-differential cross section difference 
is related to the hard function $\D \o_g$ of Eq.(\ref{d2g1fact}). 
To NLL accuracy it can be written in a factorized form as 
\beql
\lefteqn{
s^{\prime\, 2}\, 
\frac{d^2 \D \sigma_{\g^*g}(s',t_1,u_1)}{d t_1\, d u_1} \,=\, 
8\p^2\, \alpha \, \frac{x}{Q^2}\, 
\D \omega_{g}\Big(x,s_4,t_1,u_1,Q^2,m^2,\m^2,\alpha_s(\mu)\Big)\, \simeq  }\\
& & 
\nonumber
\D B^{\rm Born}_{\g^* g}(s^{\prime}, t_1, u_1)
\left\{
\d(s'+t_1+u_1) + \frac{\a_s(\m)}{\p} K^{(1)}(s',t_1,u_1)
+ \left(\frac{\a_s(\m)}{\p}\right)^2 K^{(2)}(s',t_1,u_1)
\right\}\, ,
\eeql
with the Born level hard part $\D B^{\rm Born}_{\g^*g}$ given by~\cite{Watson:1982ce,Gluck:1991in}
\beql
\label{bornfunc1pi}
\lefteqn{
\D B^{\rm Born}_{\g^*g}(s^{\prime}, t_1, u_1) \,=}\\
& & \a \, \a_s \, e_{\rm q}^2\, \p\,
\Biggl[ 
- {{t_1}\over{u_1}} - {{u_1}\over{t_1}} 
- 2m^2 \left({{1}\over{t_1}} 
+ {{1}\over{u_1}}+{{t_1}\over{u_1^2}}+{{u_1}\over{t_1^2}}\right)  
+2q^2\left({{1}\over{t_1}} 
+ {{1}\over{u_1}}+{{2}\over{s'}}\right)
\Biggr]\, \NO.
\eeql
The NLO and NNLO soft gluons corrections to NLL accuracy are 
\beql
\label{oneloopK}
\lefteqn{
K^{(1)}(s^{\prime}, t_1, u_1) \,=}\\
& & 
2\, C_A\, \left[{\ln(s_4/m^2)\over s_4}\right]_+ +\,
 \left[{1 \over s_4}\right]_+
\Biggl\{ C_A \left(
\ln\left( \frac{t_1}{u_1} \right) + {\rm{Re}}L_\b 
- \ln\left( \frac{\m^2}{m^2} \right) \right) 
- 2\, C_F \left( {\rm{Re}}L_\b + 1 \right)
\Biggr\}
\nonumber
\\
& &
+\, \d(s_4)\,  C_A \ln\left( \frac{-u_1}{m^2} \right)
\ln\left( \frac{\m^2}{m^2} \right)\, ,
\nonumber
\\[2ex]
\label{twoloopK}
\displaystyle
\lefteqn{
K^{(2)}(s^{\prime}, t_1, u_1) \,=}\\
& & 
2\, C_A^2\, \left[{\ln^3(s_4/m^2)\over s_4}\right]_+
+\, \left[{\ln^2(s_4/m^2)\over s_4}\right]_+
\Biggl\{  3\, C_A^2\,  \left( 
\ln\left( \frac{t_1}{u_1} \right) + {\rm{Re}}L_\b 
- \ln\left( \frac{\m^2}{m^2} \right)
\right) 
\NO\\
& &
- 2\, C_A 
\left( b_2 + 3\, C_F \left( {\rm{Re}}L_\b + 1 \right) \right)  \Biggr\} 
+\, \left[{\ln(s_4/m^2)\over s_4}\right]_+
\ln\left( \frac{\m^2}{m^2} \right) 
\Biggl\{  C_A^2\, \left( 
- 2 \ln\left( \frac{t_1}{u_1} \right) 
\right. 
\NO\\
& &
\left.
- 2 {\rm{Re}}L_\b + 2\ln\left({-u_1\over m^2}\right) 
+ \ln\left( \frac{\m^2}{m^2} \right)
\right) 
+ 2\, C_A\, 
\left( b_2 + 2\, C_F \left( {\rm{Re}}L_\b + 1 \right) \right)  \Biggr\}
\NO\\
& &
+\, \left[{1 \over s_4}\right]_+
\ln^2\left( \frac{\m^2}{m^2} \right) 
\left\{  - C_A^2 \ln\left({-u_1\over m^2}\right)\, 
 - \frac{1}{2}\, C_A\, b_2 \right\}\, ,\NO
\eeql
with $\mu$ the $\Bar{{\rm{MS}}}$-mass 
factorization scale, $b_2=(11 C_A - 2n_f)/12$ and 
$L_\b$ given below Eq.(\ref{softadim-res}).
We have checked that to NLL accuracy, the result 
in Eq.(\ref{oneloopK}) agrees with the exact $\co(\a_s)$ corrections 
calculated in Refs.\cite{SinNPB,Sinprep}. 

\begin{figure}[ht]
\begin{center}
\epsfig{file=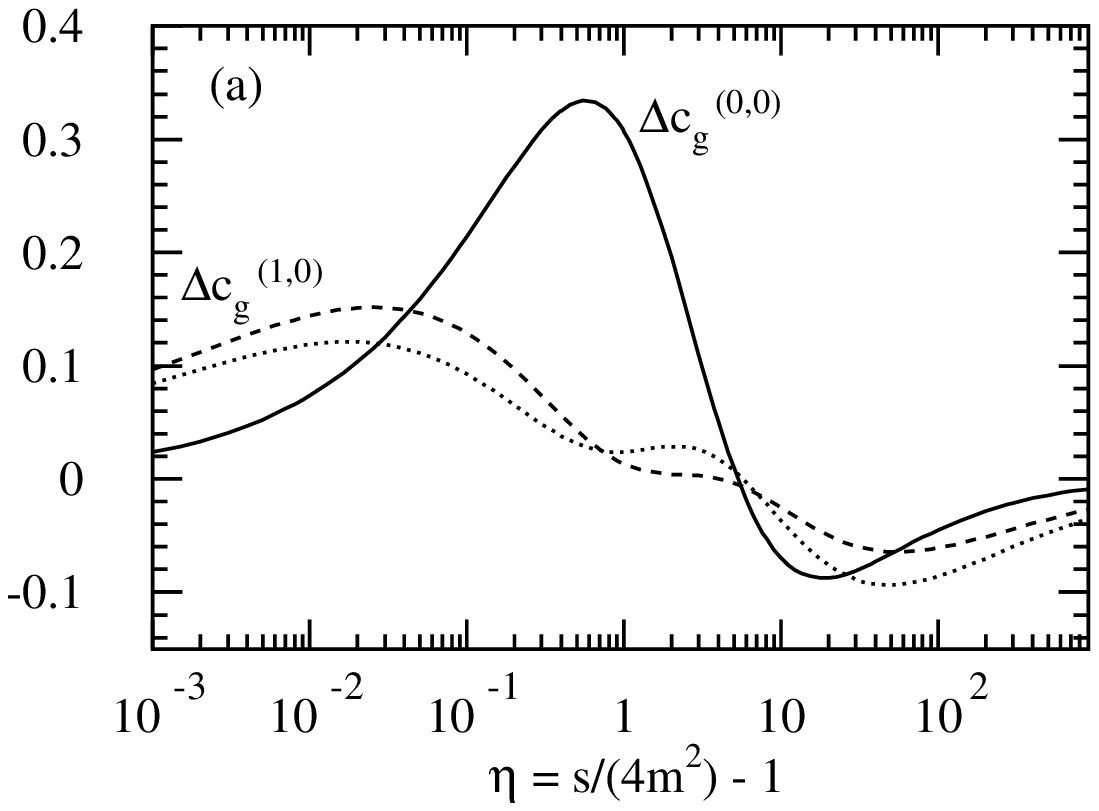,%
bbllx=50pt,bblly=120pt,bburx=285pt,bbury=460pt,angle=270,width=8.25cm}
\epsfig{file=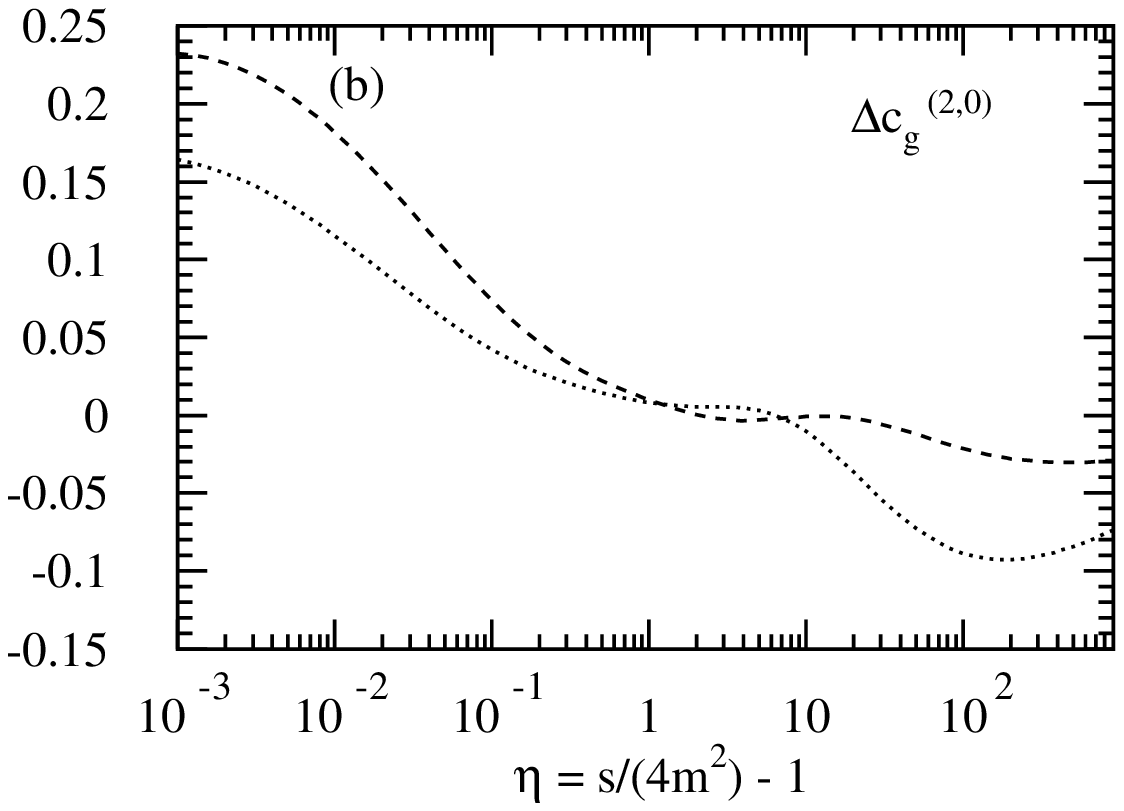,%
bbllx=50pt,bblly=120pt,bburx=285pt,bbury=460pt,angle=270,width=8.25cm}
\caption[dum]{\label{decg-fig1}{\small{
      (a)
The $\h$-dependence of the coefficient functions
$\D c^{(k,0)}_{g}(\h,\x),\;k=0,1$ 
for $Q^2=10\,{\rm GeV}^2$ with $m=1.5\,{\rm GeV}$.
Plotted are the exact result for $\D c^{(0,0)}_{g}$ (solid line) and 
the LL approximation (dotted line) and NLL approximation to $\D c^{(1,0)}_{g}$ (dashed line).
      (b) 
The $\h$-dependence of the coefficient function
$\D c^{(2,0)}_{g}(\h,\x)$ for $Q^2=10\,{\rm GeV}^2$ with $m=1.5\,{\rm GeV}$.
Plotted are the LL approximation (dotted line) and NLL approximation. 
}}}
\end{center}
\end{figure}
\begin{figure}[ht]
\begin{center}
\epsfig{file=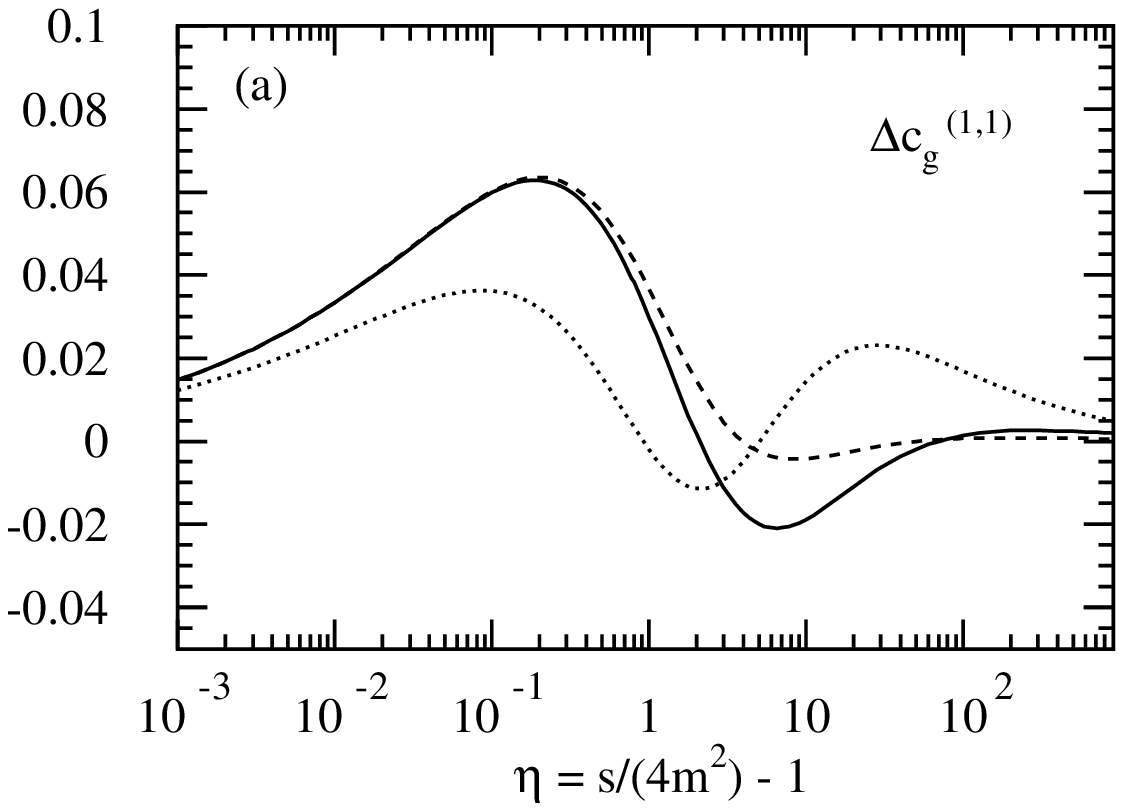,%
bbllx=50pt,bblly=120pt,bburx=285pt,bbury=460pt,angle=270,width=8.25cm}
\epsfig{file=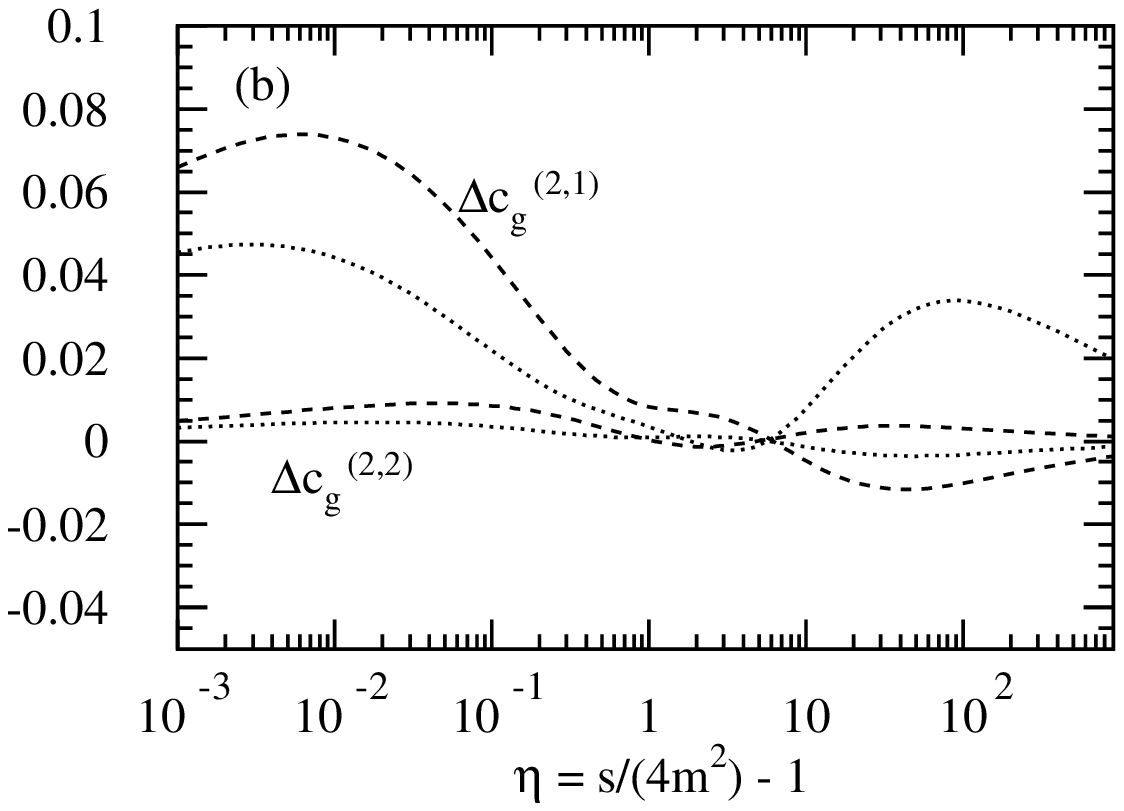,%
bbllx=50pt,bblly=120pt,bburx=285pt,bbury=460pt,angle=270,width=8.25cm}
\caption[dum]{\label{decg-fig2} {\small{
      (a) 
The $\h$-dependence of the coefficient function
$\D c^{(1,1)}_{g}(\h,\x)$ for $Q^2=10\,{\rm GeV}^2$ with $m=1.5\,{\rm GeV}$.
Plotted are the exact result (solid line), 
the LL approximation (dotted line) and the NLL approximation (dashed line).
      (b) 
The $\h$-dependence of the coefficient functions
$\D c^{(2,l)}_{g}(\h,\x),\;l=1,2$ for $Q^2=10\,{\rm GeV}^2$ with $m=1.5\,{\rm GeV}$.
Plotted are the LL approximation (dotted line) and the NLL approximation (dashed line).}}}
\end{center}
\end{figure}

Next, we perform a numerical investigation of the 
results~(\ref{oneloopK}) and (\ref{twoloopK}). 
To that end, we relate the total partonic cross section difference 
to dimensionless coefficient functions $\D c^{(k,l)}_{g}$, 
\beql
\D \s_{\g^*g}(s^{\prime},q^2,m^2) &=& \frac{\a\, \a_s\, e_{\rm q}^2}{m^2}\, 
\sum\limits_{k=0}^{\infty} (4 \p \a_s(\m))^k
\sum\limits_{l=0}^{k}
\D c^{(k,l)}_{g}(\h,\x) \ln^l\frac{\m^2}{m^2}\, .
\label{partint}
\eeql
As completely inclusive quantities the functions $\D c^{(k,l)}_{g}$  
depend only
on the scaling variables
\beql
\label{etaxidef}
\h &=&  \frac{s}{4 m^2}\, -1\, , \,\,\,\,\,\,   \,\,\,\,\,\, \,\,\,\,\,\, \,\,\,\,\,\,
\x \,=\, \frac{Q^2}{m^2}\, ,
\eeql
where $\h$ is a measure of the distance to the partonic threshold.

In Fig.\ref{decg-fig1} we plot all coefficient functions $\D c^{(k,0)}_{g}; k = 0,1,2$, 
i.e. those not accompanied by scale-dependent logarithms 
for $Q^2=10\,{\rm GeV}^2$ and $m=1.5\,{\rm GeV}$.
Only the Born function $\D c^{(0,0)}_{g}$ in Fig.\ref{decg-fig1}a is known 
exactly~\cite{Watson:1982ce,Gluck:1991in}.  
For $\D c^{(1,0)}_{g}$ in Fig.\ref{decg-fig1}a and $\D c^{(2,0)}_{g}$ in Fig.\ref{decg-fig1}b 
we can give estimates to LL and NLL accuracy. 
In the LL case, we keep only the $[\ln(s_4/m^2)/s_4]_+$ and $[\ln^3(s_4/m^2)/s_4]_+$ 
terms in Eqs.(\ref{oneloopK}) and (\ref{twoloopK}) respectively. 
We find both $\D c^{(1,0)}_{g}$ and $\D c^{(2,0)}_{g}$ to be sizable near threshold 
where the large logarithms dominate, while they tend to be numerically 
rather small at larger values of $\h$.
Moreover, we also investigated some next-to-next-to leading (NNLL) logarithmic 
contributions\footnote{\label{footnotethree}
For complete resummation to NNLL accuracy one would need to match the resummed result 
Eq.(\ref{deltaomegaresummed}) at NLO, which requires knowledge of all one-loop 
soft and virtual corrections. These are not yet available \cite{SinNPB,Sinprep}.}, 
such as the Coulomb corrections \cite{SinNPB,Sinprep,Buza:1997xr} 
for $\D c^{(1,0)}_{g}$ and for $\D c^{(2,0)}_{g}$ 
those NNLL terms, which we obtain from the expansion of the resummed result 
Eq.(\ref{deltaomegaresummed}). 
Numerically, we found these NNLL terms to have an effect 
on $\D c^{(1,0)}_{g}$ or $\D c^{(2,0)}_{g}$ of the order of $5 \%$ 
as compared to our NLL corrections.

Since no exact results for $\D c^{(1,0)}_{g}$ and $\D c^{(2,0)}_{g}$ are
available we are unable to judge the quality of our approximation.   
However, similar investigations in the case of unpolarized heavy quark production 
\cite{Laenen:1998kp}, 
where exact one-loop results are known~\cite{Laenen:1993zk},
suggest that generally NLL accuracy provides a very good approximation of the true 
result for $Q^2$ not too large\footnote{In the regime of large $Q^2/m^2$ on the other hand, 
the coefficient functions $\D c^{(k,0)}_{g}$ can be approximated with different methods 
based on the operator product expansion \cite{Buza:1997xr}.}, because $K^{(1)}$,
$K^{(2)}$ are the same as for the unpolarized case, only 
$ B^{\rm Born}_{\g^* g}$ differs from $\D B^{\rm Born}_{\g^* g}$.

Let us now turn to those coefficient functions multiplying scale-dependent logarithms.
These are $\D c^{(1,1)}_{g}$ in Fig.\ref{decg-fig2}a 
and $\D c^{(2,1)}_{g}$, $\D c^{(2,2)}_{g}$ in Fig.\ref{decg-fig2}b, which we plot 
for the same values of parameters as in Fig.\ref{decg-fig1}. 
We find in particular the estimates for $\D c^{(1,1)}_{g}$ and $\D c^{(2,1)}_{g}$ 
to NLL accuracy to be large near threshold. 
Additionally, we are able to obtain 
the exact result for the function $\D c^{(1,1)}_{g}$
by means of renormalization group methods, 
\beql
\D c^{(1,1)}_{g}\left(\h(x),\x\right)\,&=& \frac{1}{4 \p^2} 
\int\limits_{ax}^{1}\, dz\,\,
\left( b_2\, \d(1-z) - \frac{1}{2} \D P_{gg}^{(0)}(z) \right)\, 
\D c^{(0,0)}_{g}\left(\h\Bigl(\frac{x}{z}\Bigr),\x\right)\, ,
\label{ex-dc11}
\eeql
with $a=1+4m^2/Q^2$ and the one-loop splitting function $\D P_{gg}^{(0)}$ 
given in Refs.~\cite{Sasaki:1975hk,Ahmed:1976ee,Altarelli:1977zs}.
Fig.\ref{decg-fig2}a clearly shows that the approximation based on 
NLL accuracy traces the exact result (\ref{ex-dc11}) 
extremely well even at larger $\h$.
For $\D c^{(2,1)}_{g}$ we have checked that NNLL terms 
obtained from the expansion of the resummed result Eq.(\ref{deltaomegaresummed}) 
have only a small effect.

\begin{figure}[ht]
\begin{center}
\epsfig{file=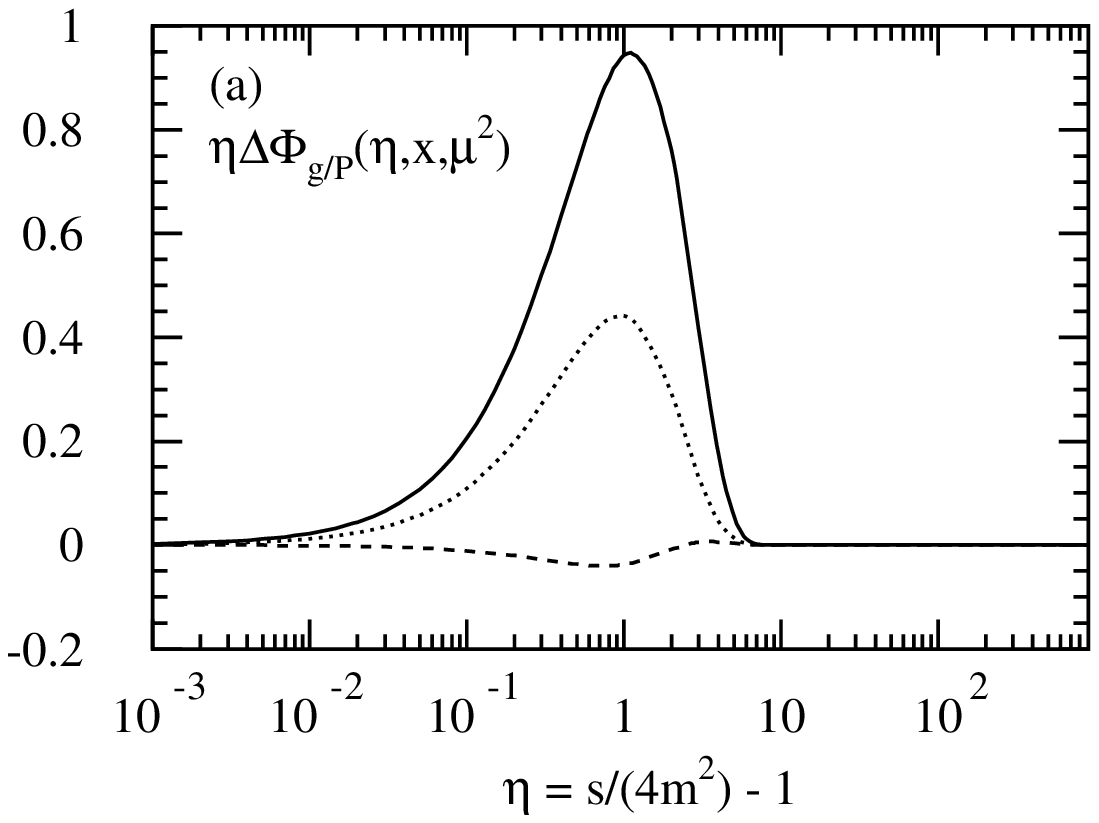,%
bbllx=50pt,bblly=120pt,bburx=285pt,bbury=460pt,angle=270,width=8.25cm}
\epsfig{file=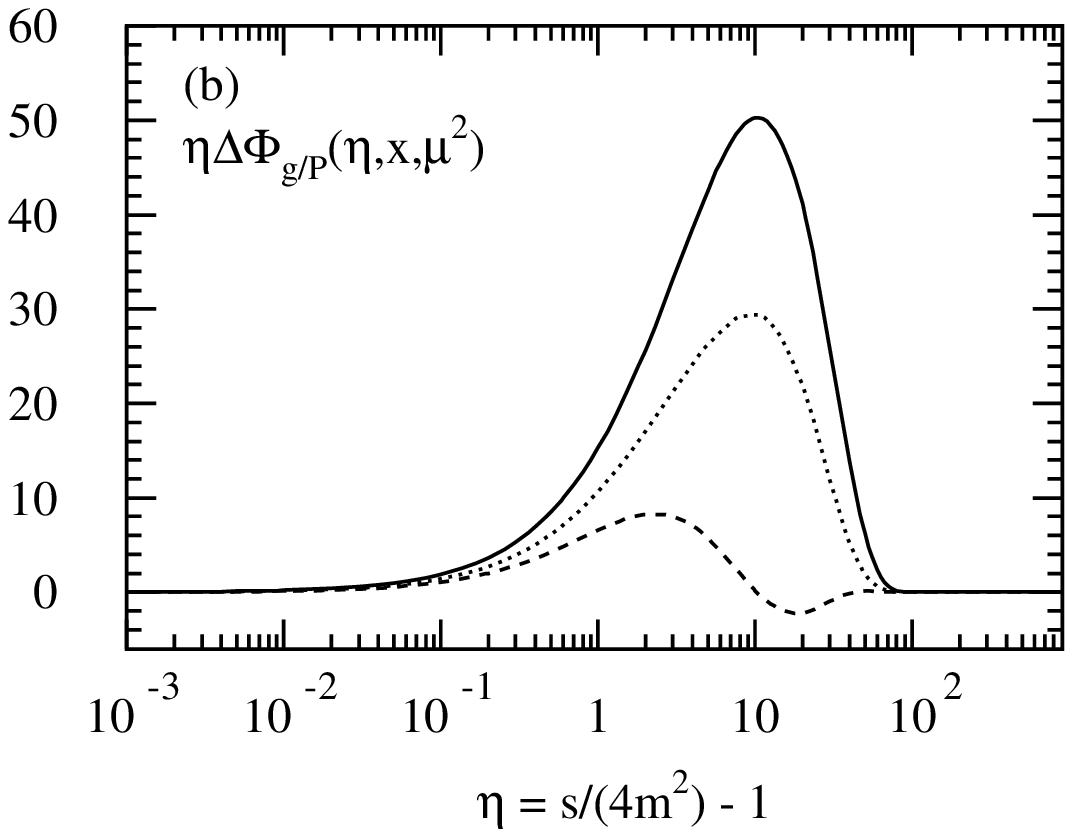,%
bbllx=50pt,bblly=120pt,bburx=285pt,bbury=460pt,angle=270,width=8.25cm}
\caption[dum]{\label{deltaphi1} {\small{
      (a) 
The polarized gluon distribution $\eta \D \f_{g/P}(\eta,x,\m^2)$ 
for $x=0.1$ and $\m = 1.5\,{\rm GeV}$. 
Plotted are the parametrizations GS~A of Ref.\cite{Gehrmann:1996ag} (solid), 
GS~C of Ref.\cite{Gehrmann:1996ag} (dashed) 
and GRSV valence of Ref.\cite{Gluck:1996yr} (dotted).
     (b) Same as (a) for $x=0.01$.
}}}
\end{center}
\end{figure}
\begin{figure}[ht]
\begin{center}
\epsfig{file=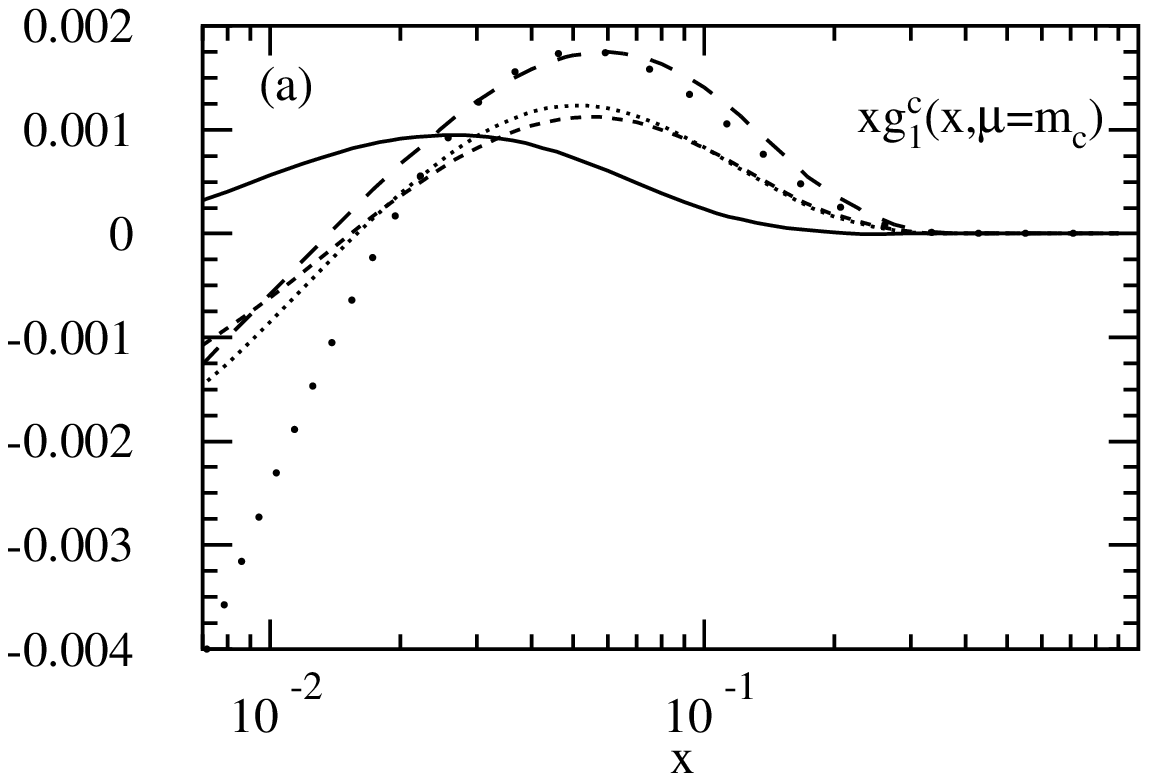,%
bbllx=50pt,bblly=110pt,bburx=285pt,bbury=450pt,angle=270,width=8.25cm}
\epsfig{file=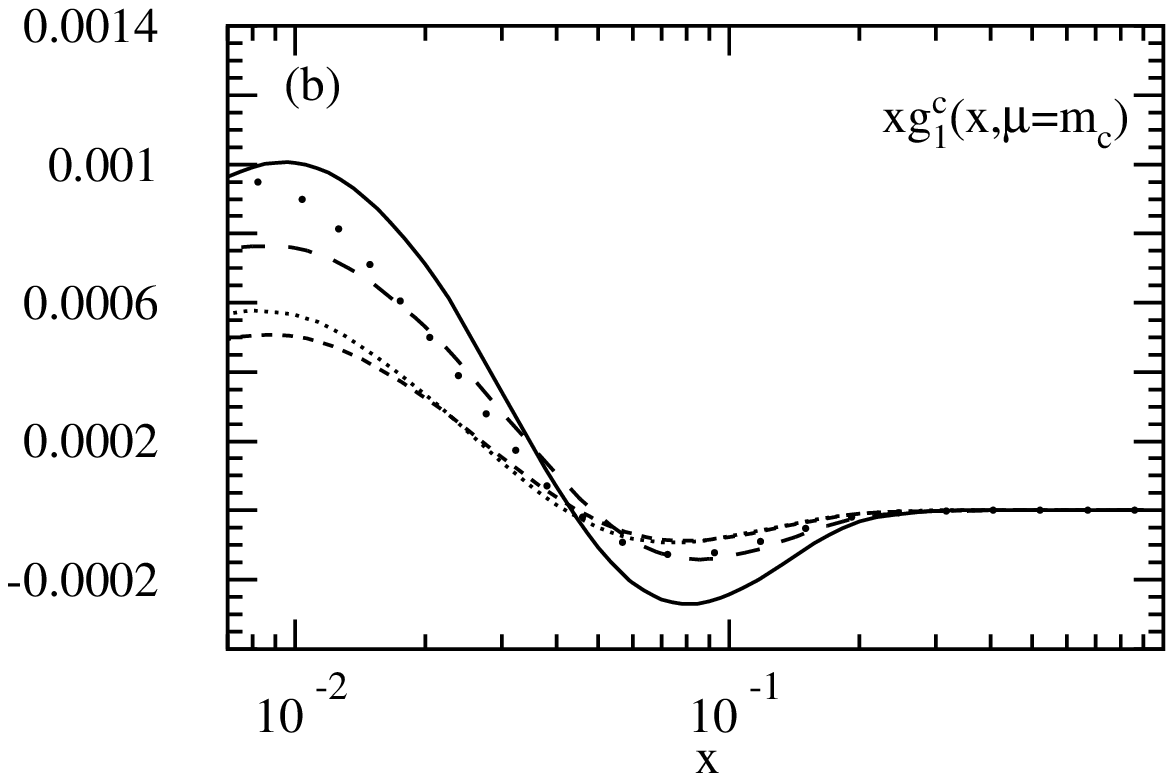,%
bbllx=50pt,bblly=110pt,bburx=285pt,bbury=450pt,angle=270,width=8.25cm}
\caption[dum]{\label{g1c-xdep1} {\small{
      (a) 
The $x$-dependence of the 
charm structure function $x g_1^{\rm c}(x,Q^2,m^2)$ 
with the gluon distribution GS~A of Ref.\cite{Gehrmann:1996ag} 
for $\m = m = 1.5\,{\rm GeV}$ and $Q^2 = 10\,{\rm GeV}^2$.
Plotted are the results at leading order (solid), 
at NLO to LL accuracy (dotted), at NLO to NLL accuracy (dashed), 
at NNLO to LL accuracy (spaced dotted) and 
at NNLO to NLL accuracy (spaced dashed).
      (b) Same as (a) with the gluon distribution GS~C of Ref.\cite{Gehrmann:1996ag}.
}}}
\end{center}
\end{figure}
\begin{figure}[ht]
\begin{center}
\epsfig{file=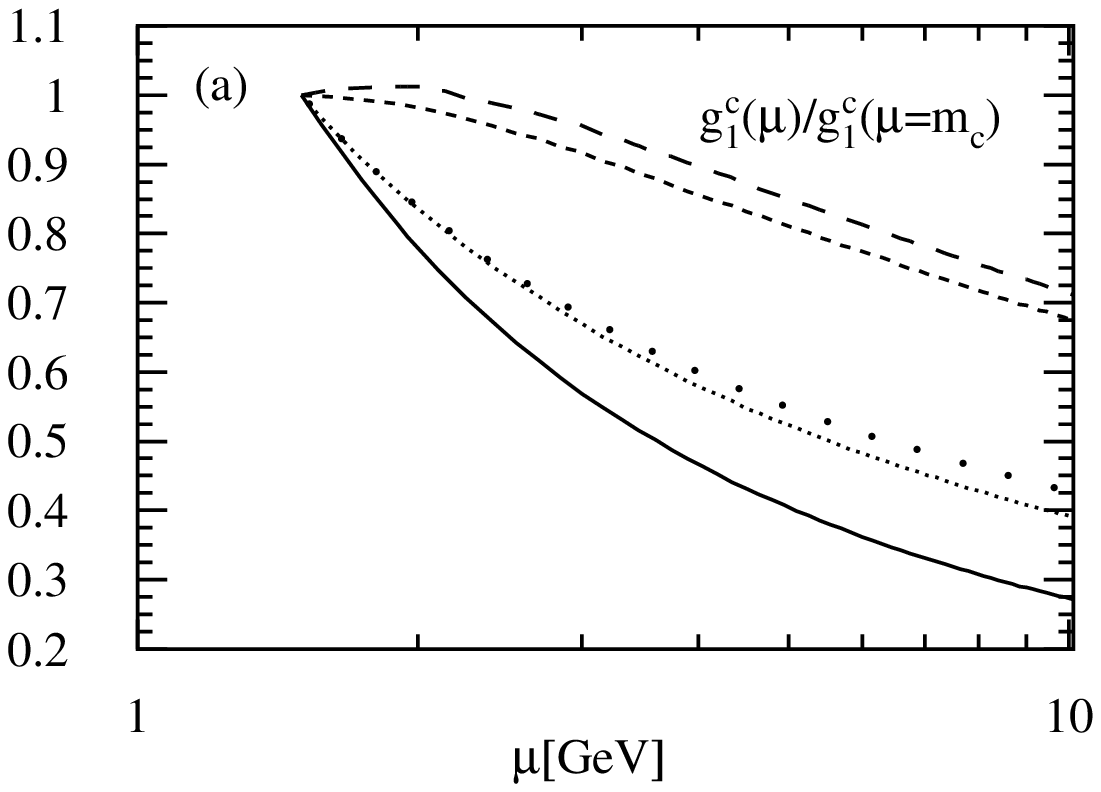,%
bbllx=50pt,bblly=120pt,bburx=285pt,bbury=460pt,angle=270,width=8.25cm}
\epsfig{file=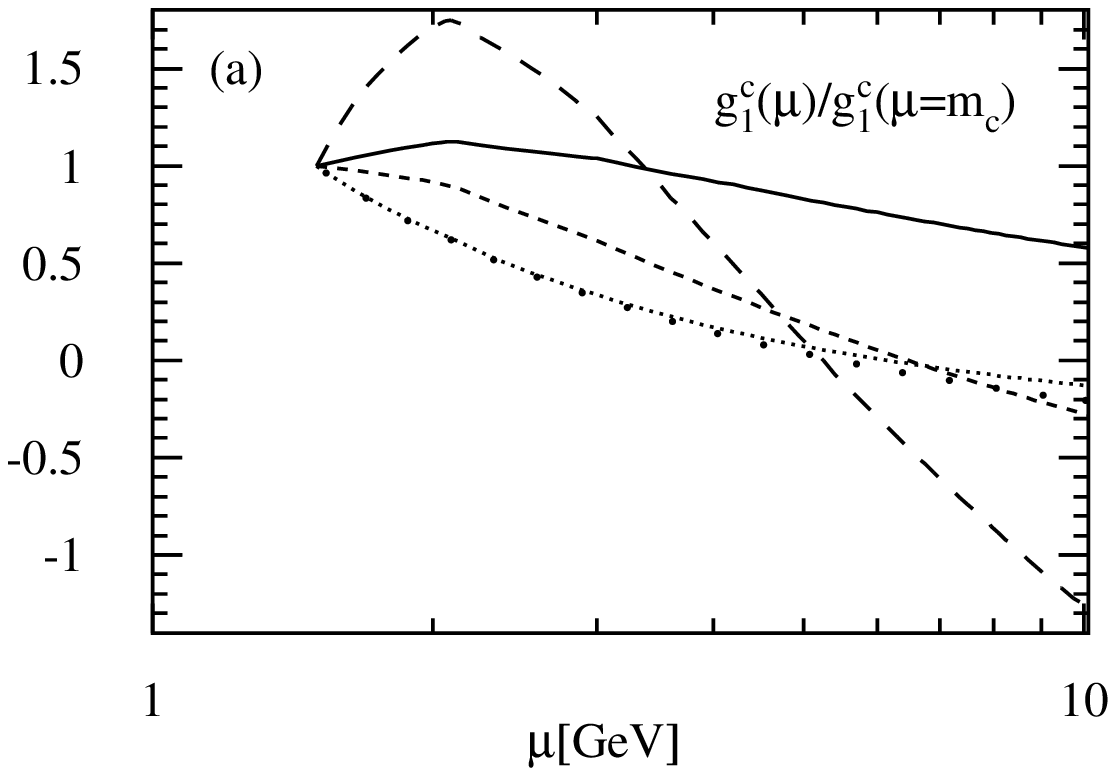,%
bbllx=50pt,bblly=120pt,bburx=285pt,bbury=460pt,angle=270,width=8.25cm}
\caption[dum]{\label{g1c-mudep1} {\small{
      (a) 
The factorization scale dependence of the 
charm structure function $g_1^{\rm c}(x,Q^2,m^2,\mu^2)/g_1^{\rm c}(x,Q^2,m^2,\mu^2=m^2)$ 
with the gluon distribution GS~A of Ref.\cite{Gehrmann:1996ag} 
for $m = 1.5\,{\rm GeV}$, $Q^2 = 10\,{\rm GeV}^2$ and $x=0.05$.
Plotted are the results at leading order (solid), 
at NLO to LL accuracy (dotted), at NLO to NLL accuracy (dashed),
at NNLO to LL accuracy (spaced dotted) and 
at NNLO to NLL accuracy (spaced dashed).
      (b) Same as (a) with the gluon distribution GS~C of Ref.\cite{Gehrmann:1996ag}.
}}}
\end{center}
\end{figure}

Let us now investigate the impact of the approximate 
higher order perturbative corrections on the inclusive hadronic structure 
function $g_1$.
In terms of coefficient functions, the charm structure function $g_{1}^{\rm c}$ 
can be expanded as follows: 
\beql
\label{charmstrucintegrated}
\lefteqn{g_1^{\rm c}(x,Q^2,m^2)} \\
&&
=\frac{\a_s\left(\mu\right) e_{{\rm c}}^2 Q^2}{8 \p^2 m^2 x}\!
\int\limits_{ax}^1 d\!z \D \f_{g/P}(z,\m^2)\,
\sum\limits_{k=0}^{\infty} (4 \p \a_s\left(\mu\right))^k
\sum\limits_{l=0}^{k}
\D c^{(k,l)}_{g}\left(x/z,\xi\right) \ln^l\frac{\m^2}{m^2}
\NO \\ 
&&=\frac{\a_s\left(\mu\right) e_{{\rm c}}^2 }{2 \p^2 }\! 
\int\limits_{-\infty}^A d\!\left(\log_{1\!0} \eta\right) 
\ln10 \ \eta \D \f_{g/P}(\eta,x,\m^2)\,
\sum\limits_{k=0}^{\infty} (4 \p \a_s\left(\mu\right))^k
\sum\limits_{l=0}^{k}
\D c^{(k,l)}_{g}\left(\h,\xi\right) \ln^l\frac{\m^2}{m^2} \ , \NO
\eeql
where $a=1+4m^2/Q^2$, $A=\log_{10}(\xi\{1/x- 1\}/4 -1)$,
and 
$\D \f_{g/P}$ represents 
the polarized gluon distribution.
We recall that the $\Bar{{\rm{MS}}}$-scheme has 
been chosen and 
that contributions from light initial state quarks
are neglected. 
For $\D \f_{g/P}$ we compare the parametrizations of Refs.\cite{Gehrmann:1996ag,Gluck:1996yr}.
For our analysis at NLO (LO) we use a 2-loop (1-loop) running
coupling with $n_f = 3$ light flavours, a charm (pole) mass of 
$m =1.5\,{\rm GeV}$~\cite{Laenen:1993zk}, 
and $\L_{\rm QCD} = 0.232~{\rm GeV}$. 

In Eq.(\ref{charmstrucintegrated}), we have chosen to rewrite 
the standard expression for the convolution of $g_1^{\rm c}$ in a form, 
that facilliates the investigation of partonic threshold effects on $g_1^{\rm c}$.
To do so, we plot $\eta \D \f_{g/P}$ as a function of $\eta$ in Fig.\ref{deltaphi1}, 
and examine the support it gives to the coefficient functions in the second expression 
of Eq.(\ref{charmstrucintegrated}).
By comparing Fig.\ref{deltaphi1} with Figs.\ref{decg-fig1} and \ref{decg-fig2} 
one can directly judge over which ranges of $\eta$ the function $\eta \D \f_{g/P}$ 
becomes large and hence samples the partonic coefficient functions.

Fig.\ref{deltaphi1} reveals that for all parametrizations this happens indeed in the 
threshold region for $x\gwig 0.01$ and at scales around $Q^2 \simeq 10\,{\rm GeV}^2$. 
We therefore expect our estimates for $\D c^{(k,l)}_{g}$ to provide 
a good description of the true higher order corrections for $g_{1}^{\rm c}$ for
$x\gwig 0.01$. 
The gluon densities from set A of Ref.\cite{Gehrmann:1996ag} and 
the valence scenario of Ref.\cite{Gluck:1996yr} are both positive and 
similar in shape, while the gluon density from set C of Ref.\cite{Gehrmann:1996ag} 
relaxes the positivity constraint on $\D \f_{g/P}$ and 
oscillates.{\footnote{Recent results from the HERMES collaboration \cite{Airapetian:1999ib} 
indicate a positive ratio of polarized over unpolarized gluon distribution 
$\D \f_{g/P}/ \f_{g/P}$ at intermediate $x$.}}

Next, we investigate the $x$-dependence of $g_1^{\rm c}$. 
In particular we are interested in a comparision of our NLO and NNLO estimates 
(\ref{oneloopK}) and (\ref{twoloopK}) with 
the leading order result~\cite{Watson:1982ce,Gluck:1991in,Stratmann:1996xy}.
In Fig.\ref{g1c-xdep1}a we plot the leading order result for $x g_1^{\rm c}$  
as well as the NLO and the NNLO approximations over a range $0.007 \le x \le 1$. 
We choose $Q^2 = 10\,{\rm GeV}^2$, a fixed value of the factorization 
scale $\m = m = 1.5\,{\rm GeV}$ and the gluon parametrization GS~A of Ref.\cite{Gehrmann:1996ag}.
We obtain similar results with the gluon distribution in  
the valence scenario of Ref.\cite{Gluck:1996yr}.
We find that the perturbative corrections are sizable, both in the region 
of intermediate $x$, around $0.05$ and at smaller $x$, where however, the 
approximation is
less certain to work well. 
To assess the quality of our approximation, we compare at each order our results 
to LL and to NLL accuracy.
At intermediate $x$, the small differences between LL and NLL accuracy show a  
very good stability of the threshold approximation for the description of $g_1^{\rm c}$.
Towards smaller $x$ however, these deviations increase and indicate that our approximate 
methods fail for $x\le 0.01$.

In Fig.\ref{g1c-xdep1}b we repeat the plot of $x g_1^{\rm c}$ for the same parameters as 
in Fig.\ref{g1c-xdep1}a but with the gluon density GS~C of Ref.\cite{Gehrmann:1996ag}.
Again, the perturbative corrections are sizable over the whole range in $x$, 
but the shape of the curves is completely different. 
Both the coefficient functions and the gluon density oscillate, which leads 
in particular at intermediate $x$ around $0.1$ to destructive interference, 
with $g_1^{\rm c}$ being only marginally different from zero. 

The final issue we study is the dependence of $g_1^{\rm c}$ on the 
mass factorization scale. 
In general, leading order calculations exhibit a strong sensitivity on the 
factorization scale, which is usually reduced once higher order 
corrections are taken into account.
In addition, there are general arguments supporting a reduction in scale
dependence from including soft gluon effects~\cite{Sterman:2000pu}.
Therefore, we are interested in the effect of our NLL-approximate NLO and NNLO 
results (\ref{oneloopK}) and (\ref{twoloopK}) on the scale dependence of $g_1^{\rm c}$ 
in comparision to the leading order calculation~\cite{Watson:1982ce,Gluck:1991in,Stratmann:1996xy}. 
Note however, that the arguments of Ref.\cite{Sterman:2000pu} leading 
to genuine NLO soft gluon resummations rely on NNLL accuracy, which is not yet available 
for $g_1^{\rm c}$, see footnote \ref{footnotethree}.

We plot the ratio $g_1^{\rm c}\left(\mu\right)/g_1^{\rm c}\left(\mu=m\right)$ 
over a range $m \le \m \le 10 {\rm GeV}$ and fix $m = 1.5\,{\rm GeV}$, 
$Q^2 = 10\,{\rm GeV}^2$ and $x=0.05$. 
In Fig.\ref{g1c-mudep1}a we use the parametrization GS~A of Ref.\cite{Gehrmann:1996ag},
but the following conclusions hold also for the gluon density in the valence scenario 
of Ref.\cite{Gluck:1996yr}.
We find at NLO that the corrections based on LL accuracy only are not sufficient to 
reduce the scale dependence. 
Clearly, soft gluon effects need to be approximated at least to NLL accuracy
to achieve the desired result,  
a feature that has also been noticed in studies of unpolarized heavy quark 
production \cite{Laenen:1998kp}.
Our best estimate for $g_1^{\rm c}$ makes use of the exact expression for the 
coefficient function $\D c^{\left(1,1\right)}$ of Eq.(\ref{ex-dc11}). 
However, since it differs only slightly from the NLO result to NLL accuracy 
for the chosen parameters in Fig.\ref{g1c-mudep1}, we do not display it here.

In Fig.\ref{g1c-mudep1}b we repeat the analysis for the parametrization GS~C 
of Ref.\cite{Gehrmann:1996ag}. 
In this case, the NLO and NNLO approximations do not reduce the scale dependence.
This is due to the oscillating gluon density, which leads to $g_1^{\rm c}$ 
being close to zero at the $x$-value chosen, 
and even causes $g_1^{\rm c}$ to change sign, depending on the scale. 

To summarize, the study of the charm structure function as in Figs.\ref{g1c-xdep1} and \ref{g1c-mudep1} 
shows that the soft gluon estimates of higher order corrections 
to the coefficient functions are well under control and give stable predictions 
for $g_1^{\rm c}$ at scales $Q^2$ not too large and $x\gwig 0.01$.
On the other hand, in the chosen kinematical range the effects of 
higher orders do not upset the
sensitivity to the polarized gluon distribution function, with 
different gluon parametrizations leading to qualitatively 
different behaviour for $g_1^{\rm c}$.
Therefore, measurements of the charm structure function allow to further
constrain 
the polarized gluon density.
Finally, we found that the sizable variation of $g_1^{\rm c}$ at leading 
order due to different values of the charm mass are not
reduced by our approximate higher order corrections. 

\section{Conclusions}
\label{conclusions}

In this letter we have investigated the effects of soft gluons 
in the case of polarized leptoproduction of heavy quarks.
Our study helps to keep theoretical control to next-to-leading 
logarithmic accuracy 
over large higher order corrections due to threshold logarithms and we 
have presented analytical results for the all-order resummed cross section 
in single-particle inclusive kinematics. 

Moreover, we have provided both analytical and numerical results for the approximate 
NLO and NNLO perturbative corrections to the deep-inelastic charm structure 
function $g_1^{\rm c}$, to NLL accuracy.
As a consequence of the scattering reaction being largely 
driven by initial state gluons with not much more energy than
required to produce the final state, we have found our 
analysis of $g_1^{\rm c}$ to be well applicable in the kinematical 
range accessible to the HERMES and COMPASS experiments.
In this region, our NLO and NNLO estimates can help to reduce 
theoretical uncertainties and may assist in the theoretical interpretation 
of future $g_1^{\rm c}$ measurements.

\subsection*{Acknowledgments}

We would like to thank E. Laenen, P. Mulders and J. Smith 
for fruitful discussions. 
We are also grateful to E. Laenen for comments on the manuscript and 
to J. Smith for providing us with Ref.\cite{Sinprep}.

This work is part of the research programme of the 
Foundation for Fundamental Research of Matter (FOM) and 
the National Organization for Scientific Research (NWO).

{\footnotesize{      

}}


\begin{thebibliography}{10}

\bibitem{Anselmino:1995gn}
M.~Anselmino, A.~Efremov, and E.~Leader,
\newblock Phys. Rept. {\bf 261}, 1 (1995), hep-ph/9501369.

\bibitem{Cheng:1996ri}
H.-Y. Cheng,
\newblock Int. J. Mod. Phys. {\bf A11}, 5109 (1996).

\bibitem{Cheng:2000xb}
H.-Y. Cheng,
\newblock (2000), hep-ph/0002157.

\bibitem{Lampe:1998eu}
B.~Lampe and E.~Reya,
\newblock (1998), hep-ph/9810270.

\bibitem{Ellis:1974kp}
J.~Ellis and R.~Jaffe,
\newblock Phys. Rev. {\bf D9}, 1444 (1974).

\bibitem{unknown:1974su}
J.~Ellis and R.~Jaffe,
\newblock Phys. Rev. {\bf D10}, 1669 (1974).

\bibitem{hermescharm}
HERMES, M.~Amarian {\em et~al.},
\newblock (1997),
\newblock HERMES 97-004.

\bibitem{compass}
COMPASS, Y.~Alexandrov {\em et~al.},
\newblock (1996),
\newblock CERN/SPSLC 96-14.

\bibitem{Kidonakis:1999ze}
N.~Kidonakis,
\newblock Int. J. Mod. Phys. {\bf A15}, 1245 (2000), hep-ph/9902484.

\bibitem{Weber:1992wd}
A.~Weber,
\newblock Nucl. Phys. {\bf B382}, 63 (1992).

\bibitem{SinNPB}
J.~Smith,
\newblock Nucl. Phys. {\bf B} (Proc. Suppl.) {\bf 79}, 602 (1999).

\bibitem{Sinprep}
J.~Smith and W.~L. van Neerven,
\newblock in preparation.

\bibitem{Bojak:1998bd}
I.~Bojak and M.~Stratmann,
\newblock Phys. Lett. {\bf B433}, 411 (1998), hep-ph/9804353.

\bibitem{Bojak:1998zm}
I.~Bojak and M.~Stratmann,
\newblock Nucl. Phys. {\bf B540}, 345 (1999), hep-ph/9807405.

\bibitem{Contogouris:2000en}
A.~P.~Contogouris, Z.~Merebashvili and G.~Grispos,
\newblock Phys.\ Lett.\  {\bf B482}, 93 (2000), hep-ph/0003204.

\bibitem{Merebashvili:2000ya}
Z.~Merebashvili, A.~P.~Contogouris and G.~Grispos,
\newblock (2000) hep-ph/0007050.

\bibitem{Schienbein:1997sb}
I.~Schienbein,
\newblock Phys. Rev. {\bf D59}, 013001 (1999), hep-ph/9711507.

\bibitem{Zijlstra:1994sh}
E.~B. Zijlstra and W.~L. van Neerven,
\newblock Nucl. Phys. {\bf B417}, 61 (1994).

\bibitem{Zijlstra:Erratum}
E.~B. Zijlstra and W.~L. van Neerven,
\newblock Nucl. Phys. {\bf B426}, 245 (1994).

\bibitem{Buza:1997xr}
M.~Buza, Y.~Matiounine, J.~Smith, and W.~L. van Neerven,
\newblock Nucl. Phys. {\bf B485}, 420 (1997), hep-ph/9608342.

\bibitem{Contopanagos:1997nh}
H.~Contopanagos, E.~Laenen, and G.~Sterman,
\newblock Nucl. Phys. {\bf B484}, 303 (1997), hep-ph/9604313.

\bibitem{Sterman:1987aj}
G.~Sterman,
\newblock Nucl. Phys. {\bf B281}, 310 (1987).

\bibitem{Kidonakis:1997gm}
N.~Kidonakis and G.~Sterman,
\newblock Nucl. Phys. {\bf B505}, 321 (1997), hep-ph/9705234.

\bibitem{Laenen:1998qw}
E.~Laenen, G.~Oderda, and G.~Sterman,
\newblock Phys. Lett. {\bf B438}, 173 (1998), hep-ph/9806467.

\bibitem{Laenen:1998kp}
E.~Laenen and S.-O. Moch,
\newblock Phys. Rev. {\bf D59}, 034027 (1999), hep-ph/9809550.

\bibitem{Collins:1981uk}
J.~C. Collins and D.~E. Soper,
\newblock Nucl. Phys. {\bf B193}, 381 (1981).

\bibitem{Mertig:1996ny}
R.~Mertig and W.~L. van Neerven,
\newblock Z. Phys. {\bf C70}, 637 (1996), hep-ph/9506451.

\bibitem{Vogelsang:1996im}
W.~Vogelsang,
\newblock Nucl. Phys. {\bf B475}, 47 (1996), hep-ph/9603366.

\bibitem{Kodaira:1982nh}
J.~Kodaira and L.~Trentadue,
\newblock Phys. Lett. {\bf 112B}, 66 (1982).

\bibitem{Watson:1982ce}
A.~D. Watson,
\newblock Zeit. Phys. {\bf C12}, 123 (1982).

\bibitem{Gluck:1991in}
M.~Gluck, E.~Reya, and W.~Vogelsang,
\newblock Nucl. Phys. {\bf B351}, 579 (1991).

\bibitem{Laenen:1993zk}
E.~Laenen, S.~Riemersma, J.~Smith, and W.~L. van Neerven,
\newblock Nucl. Phys. {\bf B392}, 162 (1993).

\bibitem{Sasaki:1975hk}
K.~Sasaki,
\newblock Prog. Theor. Phys. {\bf 54}, 1816 (1975).

\bibitem{Ahmed:1976ee}
M.~A. Ahmed and G.~G. Ross,
\newblock Nucl. Phys. {\bf B111}, 441 (1976).

\bibitem{Altarelli:1977zs}
G.~Altarelli and G.~Parisi,
\newblock Nucl. Phys. {\bf B126}, 298 (1977).

\bibitem{Gehrmann:1996ag}
T.~Gehrmann and W.~J. Stirling,
\newblock Phys. Rev. {\bf D53}, 6100 (1996), hep-ph/9512406.

\bibitem{Gluck:1996yr}
M.~Gluck, E.~Reya, M.~Stratmann, and W.~Vogelsang,
\newblock Phys. Rev. {\bf D53}, 4775 (1996), hep-ph/9508347.

\bibitem{Airapetian:1999ib}
HERMES, A.~Airapetian {\em et~al.},
\newblock Phys. Rev. Lett. {\bf 84}, 2584 (2000), hep-ex/9907020.

\bibitem{Stratmann:1996xy}
M.~Stratmann and W.~Vogelsang,
\newblock Z. Phys. {\bf C74}, 641 (1997), hep-ph/9605330.

\bibitem{Sterman:2000pu}
G.~Sterman and W.~Vogelsang,
\newblock (2000) hep-ph/0002132.

\end{thebibliography}
\end{document}